\documentclass[prb,twocolumn,superscriptaddress]{revtex4}
\usepackage{graphicx}

\begin{document}

\title{On the interpretation of the Nernst effect measurements in the
cuprates}

\author{Iddo Ussishkin}
\affiliation{William I. Fine Theoretical Physics Institute, University
of Minnesota, Minneapolis, Minnesota 55455}
\author{S. L. Sondhi}
\affiliation{Department of Physics, Princeton University, Princeton,
New Jersey 08544}

\begin{abstract}
We consider the large Nernst signal discovered by Ong and collaborators
in hole-doped cuprates, in particular in the pseudogap regime. Based on
our previous quantitative calculations together with Huse [Phys. Rev.
Lett. \textbf{89}, 287001 (2002)], we discuss the interpretation of the
experimental observations as arising from superconducting fluctuations
and its relation to the vortex scenario proposed by Ong. We also
comment on the implications of the Nernst analysis for understanding
the full range of pseudogap phenomena.
\end{abstract}

\maketitle

The Nernst effect is the appearance of a transverse electric field
$E_y$ in response to a temperature gradient $(- \nabla T) \parallel
\mathbf{\hat x}$, in the presence of a perpendicular magnetic field
$\mathbf{B} \parallel \mathbf{\hat z}$ and under open circuit
conditions (Fig.~1). Conventional wisdom holds that the effect is small
in the normal state of metals where transport by quasiparticles is
dominant, as is indeed observed, e.g., in the normal state of
conventional superconductors. In contrast, conventional wisdom also
holds that the Nernst effect is large in the superconducting state
where a new set of excitations - vortices - now enter the picture.

\begin{figure}
\includegraphics[width=3in]{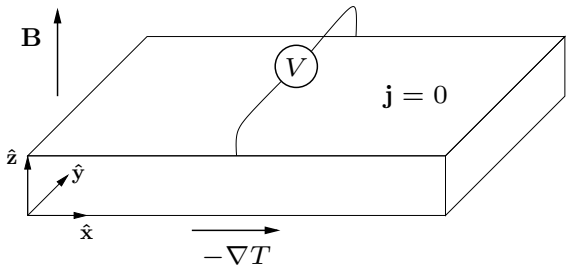}
\caption{\label{setup} Basic geometry of the Nernst experiment.}
\end{figure}

The intuition behind the wisdom is easily stated. The temperature
gradient requires that a heat current flow across the sample in the $x$
directions while the open circuit requires that there be no associated
electrical current. With quasiparticles this is accomplished by
counterflows of hot and cold electrons which then give rise to opposed
Hall voltages in the presence of the magnetic field and nearly cancel.
Once vortices are available, they can carry heat without transporting
charge and by the Josephson relation for phase slips they support a
voltage transverse to their direction of flow. This intuition is not
perfect---e.g., in particle-hole symmetric situations the two species
of quasiparticles can carry a heat current without transporting charge
and thus give Nernst signals that add, while there isn't really a
theory of the Nernst effect from vortex motion as we explain below.
Nevertheless, the expected contrast works fairly well in practice.

Recent measurements of the Nernst effect in the high-temperature
cuprate superconductors by Ong and collaborators
\cite{xu-etal,wang-etal} have challenged this conventional wisdom and
pose a problem for the theory of the Nernst effect as well as for our
evolving understanding of the cuprates. More precisely, the experiments
have uncovered a a large Nernst signal in the non-superconducting state
of hole-doped cuprates, at temperatures well above the critical
temperature $T_c$. The effect is particularly pronounced in underdoped
samples, extending well into the ``pseudogap'' region of the cuprate
phase diagram. We note that experimentally its existence was observed
by several experimental groups.~\cite{xu-etal,wang-etal,capan,wen} In
addition, there are several earlier precursor observations reported in
optimally doped samples.\cite{huebener-exp,batlogg,hagen,clayhold,hohn}

At issue then, is how to account for a large Nernst signal in the
``normal'' state of the system. The answer is presumably related to the
fact that the normal state in the cuprates is anything but normal over
much of the phase diagram---hence our use of quotation marks. Indeed,
much theoretical and experimental effort has been directed at trying to
understand the physics of the non-superconducting pseudogap region. In
this region, anomalous behavior is observed in a variety of properties
roughly below a crossover temperature scale $T^*$, including a
pseudogap in the spectrum which lends the region its name.\cite{timusk}
The Nernst effect thus presents a new challenge in this context.

The simplest proposal that can account for the Nernst data uses no
physics beyond the conventional wisdom. As the effect is large in the
presence of superconductivity and small in its absence one is led to
the conjecture that the pseudogap region contains significant
superconducting fluctuations that give rise to the observed Nernst
signal. At this level of generality, this is the interpretation
originally put forward by Ong and
collaborators.\cite{xu-etal,wang-etal} It is also in accord with an
influential set of ideas on the pseudogap that attribute its various
anomalies to fluctuating
superconductivity.\cite{randeria,emery-kivelson,others}

More specifically, Ong and collaborators have argued that the strong
Nernst signal results from vortex degrees of freedom which arise when
the amplitude of the order parameter is essentially fixed, while the
phase is allowed to fluctuate. In this picture the onset temperature of
the Nernst effect as observed in experiment, $T_{\text{onset}}$, is
taken to be as a measure of the onset of the phase fluctuation regime.
While this picture is intuitively appealing, it has not yet proven
possible to carry out computations based on it. In contrast, we and
Huse have recently performed a calculation of the contribution of
Gaussian fluctuations to the Nernst signal.\cite{ussishkin-etal} In our
work, we found that in overdoped samples of 
La$_{2-x}$Sr$_x$CuO$_4$ (LSCO), the data well above
$T_c$ quantitatively agrees with Gaussian fluctuations with a
reasonable choice of a coherence length $\xi$ and simply setting the
mean field critical temperature $T_c^{\text{MF}}$, which enters all
Gaussian computations, equal to $T_c$. In underdoped samples, the
result for Gaussian fluctuations is smaller than the observed signal if
we keep $T_c^{\text{MF}} \approx T_c$ (and use the same $\xi$). It was
argued that a qualitative understanding may be achieved by assuming
that $T_c^{\text{MF}}$ is higher than the actual $T_c$ in the
underdoped regime due to strong fluctuations. Altogether we believe
that this work provides strong support to the interpretation of the
Nernst experiments as the work of superconducting fluctuations. As the
previous discussion was rather compressed and there are several
delicate points, in this article we will provide a fuller discussion of
the issues involved. Specifically, we will (1) clarify some aspects of
our analysis, (2) discuss its relation to the vortex interpretation,
and (3) attempt to bound what we feel the Nernst effect is telling us
regarding the physics of the pseudogap. We note that the Nernst effect
was also considered by several other groups along similar or related lines.~\cite{chicago,kontani,mukerjee,honerkamp-lee}

Before proceeding, we should note that there is still considerable
debate on the physics of the pseudogap region. While it is beyond our
brief here to attempt a survey of all the work on this problem, we note
that the ideas around range from RVB (resonating valence bond)
proposals built on spin-charge separation,\cite{pwa87,lee} through fluctuating charge
and spin order ideas,\cite{kivelson,sachdev}
to the DDW ($d$-density wave) proposal in which
the pseudogap is a phase with a distinct broken symmetry.\cite{chakravarty}
However there
is not, at present, a viable alternative scenario for the Nernst
experiments based on any of these proposals which is distinct from
invoking superconducting fluctuations. For example, recently Oganesyan
and one of us have studied the contribution of quasiparticles to the
Nernst signal, in particular in connection with the DDW scenario for
the pseudogap region and failed to find a significant
effect.\cite{oganesyan-etal} This computation is of interest as similar
computations have supported the plausibility of the DDW idea in
accounting for the photoemission experiments in the pseudogap region.

We now turn to a telegraphic discussion of our main points.

1. \emph{Order parameter dynamics:} We assume that the dynamics of the
superconducting order parameter in the regime of interest is classical.
This is likely a safe assumption above and near $T_c$ provided the
system is not too underdoped. This allows us to write a {\it
stochastic} time-dependent Ginzburg-Landau equation for the evolution
of the order parameter $\psi$,
$$
\partial_t \psi = - \Gamma {\delta F \over \delta \psi} + \zeta
$$
where $F[\psi] = |\nabla \psi|^2 + a |\psi|^2 + b |\psi|^4 + \ldots$ is
the Landau-Ginzburg-Wilson functional.~\cite{cutoff} This is Model A
for critical dynamics in Hohenberg and Halperin's
classification\cite{hohenberg-halperin} and assumes that the slowest
dynamics is that of the order parameter.

It is important to note that this equation already contains in it
\emph{all} of the physics invoked by Emery and Kivelson, by Ong and
collaborators and by ourselves. Close to $T_c$ there is a region where
phase fluctuations dominate. In this region it is possible, at least in
principle, to discuss the physics in terms of the vortex degrees of
freedom. In $d = 2$ this is rigorously so near the Kosterlitz-Thouless
transition. In $d = 3$ while there are discussions of the phase
transition in terms of vortex loops \cite{shenoy} these are much more
complicated and not nearly as useful. As the temperature is increased,
the density of the vortices increases and amplitude fluctuations become
important in addition to the phase fluctuations.\cite{fn-phaseflucs} As
the temperature is increased further, another approximation becomes
available, namely to neglect the quartic term and consider Gaussian
fluctuations. The temperatures in this region are above
$T_c^{\text{MF}}$, the temperature at which the coefficient of
$|\psi|^2$ in the free energy vanishes. Both amplitude and phase
fluctuations are important in this region, and in a vortex picture
these may be thought of as arising from a dense soup of vortices. [We
note that this discussion holds for BCS superconductors as well. In
this case, the difference between $T_c$ and $T_c^{\text{MF}}$ is
$\mathcal{O} (\Delta / E_F)$, as given by the Brout criterion. Critical
behavior is observed in a much narrower region, given by the Ginzburg
criterion as $\mathcal{O} (\Delta / E_F)^4$.]

So nothing fundamental is at issue between invoking vortices and our
work---there is a single curve for the Nernst coefficient for which we
have calculated the high temperature limit.~\cite{full} If we had
failed to find a significant answer, it would have been hard to credit
superconducting fluctuations with very much effect. Of course this is
only one quantity---one will need a similar convergence between theory
and experiment for some other to put the scenario on a really firm
footing.

2. \emph{A few caveats:} Before we continue, we should consider the
limitations of our analysis. As already said, we assume classical
fluctuations, and while this is always true asymptotically close to
$T_c$, using it over a broad range of temperatures does require the
dynamics to be more robustly classical.

In our treatment we consider the dynamics of the order parameter alone,
and neglect all other degrees of freedom. Consequently quasiparticle
contributions are ignored. Although this can be justified in BCS
(including $d$-wave) superconductors, for cuprates the microscopic
picture is not yet settled, and this assumption cannot be rigorously
justified. We also neglect the possibility that other fields may
acquire a slow dynamics. Here, the main candidate appears to be the
energy field. Including it would lead to model C (but we note that
models A and C yield the same result for the Nernst signal in the
Gaussian approximation). In these systems, slow dynamics may also arise
due to other types of incipient ordering (e.g., stripes,
antiferromagnetic order) that may in principle also affect the Nernst
signal.

As for the choice of dynamics for the order parameter, we make the
conventional assumption that all parameters are
temperature-independent, except for the coefficient of the quadratic
term which varies as $T - T_c^{\text{MF}}$. In reality, these
coefficients may acquire a non-trivial temperature dependence, which
depends on the underlying microscopics. We also assume particle-hole
symmetry (or purely relaxational dynamics for the order parameter).
Breaking of this symmetry may lead to a modification of our results.

Having stated these various caveats, we argue that nevertheless the
analysis of the model A does provide important insight for the
discussion of the Nernst experiments.

3. \emph{On the vortex ``transport entropy'':} In the past, when
describing the Nernst effect in terms of vortices on a phenomenological
level, it has been customary to identify the ``transport entropy'' of a
vortex, $S_\phi$.\cite{huebener} In the Nernst setup the temperature
gradient then applies a force $\mathbf{f} = - S_\varphi \nabla T$ on
the vortices. In this picture, the ``transport entropy'' also describes
the heat carried by a vortex. Based on these definitions, the
``transport entropy'' of the vortex may be extracted from a Nernst
measurement or calculation. However, although it has often been
suggested in the literature that this quantity is related to the
entropy of the vortex core, its relation to vortex properties remains
to be clarified.

In this context, the calculation of Caroli and Maki\cite{caroli-maki}
for the ``transport entropy'' has often been cited in the literature.
They considered the drift motion of the vortex lattice in the ordered
state at the mean field level (see also Troy and
Dorsey\cite{troy-dorsey}). As we will discuss in detail elsewhere, the
original calculation has to be corrected (so that the microscopic
results for the heat current indeed match those of the TDGL near $T_c$)
and modified to properly account for magnetization currents (see
Ref.~\onlinecite{cooper}), and the correct result is rather surprising
at first sight: The Nernst signal (and hence the ``transport entropy'')
vanishes in this approach! This is in spite of the fact that there is a
free energy cost associated with the vortex core in the model. At a
hand-waving level, this result may be understood by noting that (1) the
mean field equation is effectively at zero temperature, and (2) the
solution involves an order parameter configuration drifting in its
ground state and carrying no entropy of its own.

The bottom line of this discussion is that while the Nernst signal may
be described in terms of vortices moving along the temperature gradient
and creating electric field by phase slips, performing a calculation in
this framework is non-trivial. In particular, the ``transport entropy''
is not obviously related to the properties of the vortex 
core,\cite{fn-config} and it is not clear if it has any useful meaning
beyond its relation to the transport properties of the vortex state.

4. \emph{Gaussian results:} Above $T_c^{\text{MF}}$, the Gaussian
approximation becomes available and is analytically tractable. We
briefly describe the results of our calculation in this regime.
We first note that theoretically we calculate the conductivity and
thermoelectric tensors, $\sigma$ and $\alpha$, and not the Nernst
effect directly. The Nernst signal is then related to these linear
response coefficients by
\begin{equation}\label{TDGL}
\frac{E_y}{(-\nabla T)_x} = \frac{\alpha_{xy} \sigma_{xx} - \alpha_{xx}
\sigma_{xy}}{\sigma_{xx}^2 + \sigma_{xy}^2} \approx
\frac{\alpha_{xy}}{\sigma_{xx}} .
\end{equation}
The last expression is an approximation that can often be justified in
the present context. We note that the behavior of the Nernst signal is
dominated by the behavior of the transverse thermoelectric response
$\alpha_{xy}$, making it the quantity of interest in our discussion. We
also note that to observe an appreciable Nernst signal requires
$\sigma_{xx}$ to be small. This is one of the main reasons the
contribution of superconducting fluctuations to the Nernst signal is
not observed in conventional low-temperature superconductors.

For the contribution of superconducting fluctuations to $\alpha_{xy}$
(above $T_c$ and to linear order in $B$) we obtain\cite{ussishkin-etal}
\begin{equation}\label{alpha}
\alpha_{xy}^{\text{SC}} = \frac{1}{6 \pi} \, \frac{e}{\hbar}
\frac{\xi_{ab}^2}{\ell_B^2 s} \, \frac{1}{\sqrt{1 + (2 \xi_c / s)^2}}
\, .
\end{equation}
Here, $\xi_{ab}$ ($\xi_c$) is the temperature-dependent coherence
length parallel (perpendicular) to the layers, $s$ is the interlayer
spacing, and $\ell_B$ the magnetic length. Note that the result
essentially depends on one quantity, the coherence length, making this
result particularly suitable for comparison with experiment.

In the Gaussian approximation, the coherence length diverges at
$T_c^{\text{MF}}$, $\xi_{ab / c} = \xi_{ab / c} (0)
\sqrt{T_c^{\text{MF}} / (T - T_c^{\text{MF}})}$. In reality,
$\alpha_{xy}$ does not diverge at $T_c^{\text{MF}}$ due to fluctuations
beyond the Gaussian order which work to suppress $T_c$. The
self-consistent Hartree approximation is perhaps the simplest way to go
beyond the Gaussian approximation, and should work to some extent as
the temperature is lowered below $T_c^{\text{MF}}$. In this
approximation, Eq.~(\ref{alpha}) remains valid, but the coherence
lengths are renormalized from their Gaussian value to a lower value,
diverging at~$T_c$.

We digress, briefly, to mention that our Gaussian result also arises,
with appropriate identification of parameters, as the Aslamazov-Larkin
contribution in the microscopic theory of fluctuations about the BCS
theory. In the same microscopic theory there are other diagrams, rather
a lot of them really, that must be considered. It has been shown by one
of us that all other contributions are subdominant near
$T_c$.\cite{iu-micro} For the cuprates the applicability of such
microscopic computations is unclear which is why we prefer our current
critical dynamics perspective. Finally, we also mention that the results of the Gaussian
calculation may also be extended to finite magnetic field.\cite{unpublished}

5. \emph{Overdoped LSCO,} $T_c^{\text{MF}}$, \emph{and}
$T_{\text{onset}}$: Assuming the coherence lengths are known, the
Gaussian approximation may be used to estimate $T_c^{\text{MF}}$.
Indeed, by using reasonable numbers for the coherence lengths in
overdoped LSCO and assuming $T_c^{\text{MF}} \approx T_c$, we found
that the Gaussian approximation gives quantitatively good agreement
with the measured signal.\cite{ussishkin-etal}
This is a clear indication that
$T_c^{\text{MF}}$ cannot be much higher than $T_c$ (if it was, Gaussian
fluctuations would predict a Nernst signal at high temperatures which
is larger than the observed signal). At the same time, the onset
temperature $T_{\text{onset}}$ is much higher. The value of
$T_{\text{onset}}$ is determined by the magnitude of the fluctuation
signal and the ability of the experiment to detect it above the normal
state background. In the present case the onset temperature is high for
the same reasons the Nernst effect provides such a useful probe to
superconducting fluctuations, namely that it represents a signature of
superconductivity in a property which is very small in the normal
state. In particular, we note that $T_{\text{onset}}$ occurs well
inside the region of Gaussian fluctuations.

For overdoped LSCO we thus find that in most of the temperature range
above $T_c$ the Nernst signal is a result of Gaussian fluctuations.
Close to $T_c$ one has the region where phase fluctuations dominate
over amplitude fluctuations. While the upper boundary of this region
remains somewhat ambiguously defined, it is clear from this discussion
that it cannot be taken to be $T_{\text{onset}}$. Rather, together with
$T_c^{\text{MF}}$, it will be much closer to $T_c$. Indeed, at $x =
0.20$ we have $T_c = 28\,$K, $T_{\text{onset}} \approx 75\,$K, and the
comparison of the data with our calculations suggests a
$T_c^{\text{MF}}$ of about $35\,$K or lower.~\cite{numerics} To
conclude, the region where phase fluctuations dominate is a small
region in the vicinity of $T_c$ despite the large range of temperatures
in which a Nernst signal is observed.

\begin{figure}
\includegraphics[width=3in]{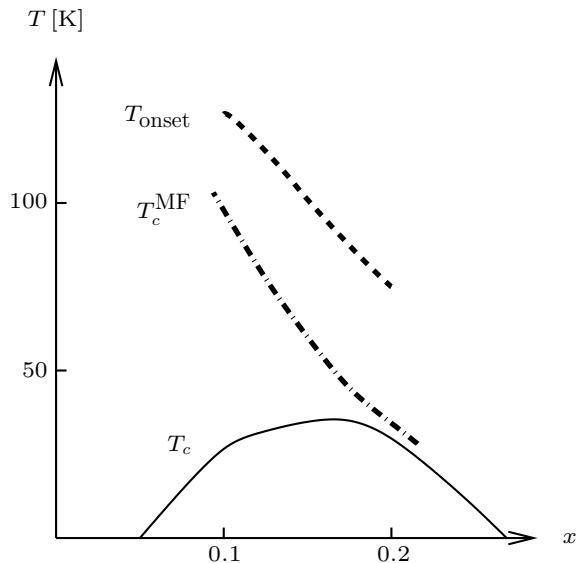}
\caption{\label{fig_TcMF} Onset temperature of the Nernst effect,
$T_{\text{onset}}$, and a suggestion of what the mean field critical
temperature $T_c^{\text{MF}}$ may look like in LSCO, as a function of
doping. The onset data is based on the measurements of Wang \emph{et
al.}~\cite{underdoped}}
\end{figure}

6. \emph{Underdoped LSCO:} What happens when one considers underdoped
samples? As we have pointed out in our paper,\cite{ussishkin-etal}
the observed signal is
significantly larger than the result for Gaussian fluctuations if
$T_c^{\text{MF}} \approx T_c$ using reasonable parameters. This implies
that a significant regime of non-Gaussian fluctuations must be included
if one is to make sense of the experiment in terms of superconducting
fluctuations. The interpretation is then that underdoping increases the
strength of the fluctuations while increasing $T_c^{\text{MF}}$, so
that a systematically large gap between $T_c^{\text{MF}}$ and $T_c$ is
opened (see Fig.~\ref{fig_TcMF}). 
As this happens, a large Nernst signal is expected from our
analysis. We note that a similar doping dependence of $T_c^{\text{MF}}$ and
$T_c$ is at the heart of discussions of the pseudogap in terms of
superconducting fluctuations;  most notably in Emery and Kivelson's
empirically deduced phase-crossover diagram for the
cuprates.\cite{emery-kivelson}

There is however one more effect which may occur with underdoping,
namely a change of the coherence length. The Nernst effect measurements
at high fields performed by Wang \emph{et al.}\cite{wang} suggest that the
coherence length decreases as the sample becomes underdoped (however,
to substantiate this result requires a theoretical understanding of the
Nernst signal in this regime). Here, we note that decreasing the
coherence length with underdoping has the effect of increasing both the
range of magnetic fields in which fluctuation effects are observable
(since $H_{c2}$ becomes higher), and the effect of strong fluctuations
(i.e., the separation between $T_c$ and $T_c^{\text{MF}}$). These
results are at the heart of the usual statement that a short coherence
length leads to a large region for the observation of fluctuation
phenomena in the cuprates. On the other hand, a shorter coherence
length implies a weaker Gaussian contribution to $\alpha_{xy}$. On
comparison with the usual implications of a short coherence, this
result is quite counterintuitive! This suggests that as the sample goes
underdoped, $T_c^{\text{MF}}$ approaches $T_{\text{onset}}$ as the
region of observability of Gaussian fluctuations decreases. One should
remember, however, that the arguments regarding the position of
$T_c^{\text{MF}}$ in the underdoped regime are of the hand-waving type,
and are not supported by a quantitative fit at this stage.

7. \emph{The ``pseudogap'':}
One of the main interests in the Nernst experiments arises from the 
possible implications for the physics of the pseudogap regime in the cuprates.
Clearly, our analysis suggests that strong superconducting fluctuations 
exist in underdoped cuprates.
One possibility, is to assume that all pseudogap phenomena arise 
somehow due to the same reason.
Alternatively, it has been suggested that the pseudogap arises due to a 
spin gap, or one of several
possible suggestions for a competing order. Or it could be that both 
piece of physics are operative in the pseudogap region.
Here, we consider what can be said about this issue from the 
perspective of the Nernst effect.

The pseudogap region is observed in a wide variety of experimental probes 
(including NMR, various resistivity measurements, and heat capacity), 
roughly below a temperature scale $T^*$.\cite{timusk}
We therefore would like to ask if (and how) $T^*$ is related to $T_c^{\text{MF}}$.
The pseudogap temperature $T^*$ is itself a
crossover temperature scale, so it is not well defined. Still, in
experiments it is typically higher than $T_{\text{onset}}$ (in LSCO, it
is roughly linear in doping, decreasing from about $350\,$K at $x =
0.08$ to about $90\,$K at $x = 0.22$, as derived from heat capacity
measurements). In particular, it is sufficiently higher than
$T_c^{\text{MF}}$ to suggest that $T^*$ and $T_c^{\text{MF}}$ are two
distinct temperature scales. Of course, different measurements may
yield different results for a crossover scale. On the other hand,
because the Nernst effect is probably the most sensitive available
probe to superconducting fluctuations, as an empirical matter one would
expect that it yields a higher crossover scale arising from
superconducting fluctuations than in other quantities.
These observations suggest that $T^*$ is not set by superconducting fluctuations 
but is of a different physical origin.
In particular, it may well be related to a spin gap or a competing order 
as suggested in the literature. However, the Nernst effect does suggest 
that superconducting fluctuations are an
important part of the story at lower temperatures, and must be incorporated in the theoretical description of the pseudogap regime.

8. \emph{Other fluctuation phenomena:} If the Nernst effect in the cuprates
indeed arises from superconducting fluctuations, the question arises whether
similar effects can be observed in other measurements. On the one hand,
there are several measurements supporting the picture of strong fluctuations.
These include ac conductivity\cite{corson} and thermodynamic measurements.\cite{meingast}
On the other hand, the dc conductivity and magnetization do not show an
enhancement which can immediately be credited to strong superconducting fluctuations.
It is therefore important to check whether these observations may be reconciled with our
interpretation of the Nernst effect.

The contribution of superconducting fluctuations to the magnetization and
conductivity (at the Gaussian level) are well known.\cite{larkin}
For the magnetization, such contribution is expected to be very small and difficult
to detect in the present case, except very close to $T_c$. 
With the conductivity the situation is different, i.e., the usual Aslamazov-Larkin 
result for the conductivity does not give a very small contribution 
already for Gaussian fluctuations. 
However, it must be noted that this result is invariably based on 
BCS microscopics, in particular, on the BCS value for the order
parameter relaxation rate~$\Gamma$.
The fluctuation contribution to the 
conductivity is inversely proportional to $\Gamma$ 
(while the Nernst is independent of it). 
Hence, it is certainly possible to reconcile the absence of large 
fluctuation contributions to the  conductivity in experiment with
our interpretation if $\Gamma$ is substantially larger then its BCS value. 
At present there is no microscopic calculation supporting this; 
however, one possibility which comes to mind is that
competing orders in the pseudogap phase provide a mechanism 
for a fast relaxation of the superconducting order parameter 
(see also Ref.~\onlinecite{honerkamp-lee}).

In conclusion, in this paper we discussed the Nernst effect in 
the cuprate and its interpretation in
terms of superconducting fluctuations. Our discussion is based 
on our previously published
quantitative calculation.\cite{ussishkin-etal} Already at the level of 
Gaussian fluctuations our
results find substantial superconducting fluctuations. Strong fluctuations, 
with a substantial separation between 
$T_c^{\text{MF}}$ and $T_c$, occur in this region. On 
the other hand, $T_c^{\text{MF}}$ is significantly lower than $T^*$, suggesting a different 
source for other pseudogap phenomena in the upper reaches of the pseudogap region.

We thank Steve Kivelson, Tom Lubensky, Subroto Mukerjee, Phuan Ong, 
Yayu Wang, and especially David Huse for enlightning conversations and 
their valuable input. This work was supported by the National Science 
Foundation through grants EIA-02-10736 and DMR-02-13706 and by the 
David and Lucile Packard Foundation.


\begin{thebibliography}{22}
\bibitem{others}
The notion that fluctuations are important in the physics of
superconductivity at low dopings has appeared in many guises in the
work of many people. While a proper accounting of its history would be
more appropriate to a full-scale review of the topic, here we mention
the early contributions of Anderson, who referred to the importance of
phase fluctuations in Ref.~\onlinecite{pwa87}, the Uemura plot\cite{uemura}
which pointed to the relationship between the superfluid
density and $T_c$ later formalized by Emery and Kivelson and the paper
by Doniach and Inui.\cite{doniach}
\bibitem{cutoff}
We note that in this model there is also a dependence on the
ultraviolet cutoff and the way that it is implemented.
\bibitem{fn-phaseflucs}
We are indulging in somewhat standard abuse of language here. Vortices
generically involve amplitude fluctuations since the order parameter
will be suppressed inside the core. So the increased amplitude
fluctuations refer to variations of the amplitude beyond this
``minimum''.
\bibitem{full}
The full curve was recently analyzed numerically by Mukerjee and Huse
at finite magnetic field;\cite{mukerjee} their result agrees with the
Gaussian computation at high temperatures, as expected.
\bibitem{fn-config}
Indeed Mukerjee and Huse \cite{mukerjee} have given evidence that the
entropy transported is configurational.
\bibitem{numerics}
There is also support to this picture from the numerical simulations of
Mukerjee and Huse,\cite{mukerjee} who fit data for this sample below
$T_c$ at finite magnetic field with two-dimensional simulation results
using $T_c^{\text{MF}} = 39\,$K.
\end{thebibliography}
\end{document}